\documentclass[copyright,creativecommons]{eptcs}
\providecommand{\event}{ACL2 Workshop 2025}

\usepackage{iftex}
\usepackage{amsmath}
\usepackage{tikz}
\usetikzlibrary{arrows.meta,decorations.pathmorphing,positioning,fit,decorations.pathreplacing,bending,calc}
\usepackage{alltt}
\usepackage{syntax}

\usepackage[T1]{fontenc}
\usepackage{upquote}
\usepackage{etoolbox} 

\robustify{\texttt}

\ifpdf
  \usepackage{underscore}         
  \usepackage[T1]{fontenc}        
\else
  \usepackage{breakurl}           
\fi

\newtheorem{theorem}{Theorem}

\title{On Automating Proofs of Multiplier Adder Trees \\using the RTL Books}
\author{Mayank Manjrekar
\institute{Austin Design Center\\ Arm Inc.}
\email{mayank.manjrekar2@arm.com}
}
\def\titlerunning{On Automating Proofs of Multiplier Adder Trees using the RTL Books}
\def\authorrunning{M. Manjrekar}
\begin{document}
\maketitle

\newcommand{\pp}{\mathit{pp}}

\begin{abstract}
  We present an experimental, verified clause processor
  \texttt{ctv-cp} that fits into the framework used at Arm for formal
  verification of arithmetic hardware designs. This largely automates
  the ACL2 proof development effort for integer multiplier modules
  that exist in designs ranging from floating-point division to matrix
  multiplication.
\end{abstract}

\section{Introduction}
Formal verification of multipliers is a difficult problem. At Arm, we
have a well-established methodology~\cite{cma2022,russinoff2022formal}
for verifying arithmetic hardware designs. Verification of a design is
a two-step process. First, we model the RTL using the RAC programming
language~\cite{russinoff2022formal}, a restricted subset of C++
augmented with AC datatypes~\cite{acdatatypes}, and prove it
equivalent to the design using an industrial equivalence checker.
Second, we use the RAC parser to automatically translate the RAC model
into ACL2 and prove that it is correct with respect to a high-level
specification; we use mathematical abstractions in the RTL
library~\cite{rtlbooks} where, e.g., floating-point operations are
specified using rational numbers. Developing the RAC model requires a
delicate balance: a higher level of abstraction favors ACL2 proofs but
a lower level favors equivalence checks. In this paper, we present an
experimental, verified clause processor \texttt{ctv-cp}
\cite{ctv-cp-link} that fits into our framework and largely automates
the ACL2 proof development effort for integer multipliers. It allows
the RAC model to directly mimic a large portion of the RTL, thereby
simplifying model development and facilitating fast equivalence
checks.

The design of an integer multiplier may be divided into two parts: the
generation and the summation of partial products. Various optimization
techniques are employed for performance, but the above partitioning is
accurate in principle. Summation of the partial products is done by a
{\em compression tree} circuit that has the largest proportion of the
multiplier's area. The compression tree performs a sequence of steps
to eventually reduce the number of partial products to two. The two
output vectors of the reduction are added together using a
carry-propagate adder. Each reduction step is typically implemented
using a 3:2 compressor, whose output vectors, \emph{sum} and
\emph{carry}, have the following formula:
\(sum = x \oplus y \oplus z,\ carry = (x \wedge y) \vee (x \wedge z) \vee (y \wedge z)\).
Figure \ref{fig:8x8-bit-matrix} shows a bit-matrix representation of
a compression tree of a simple \(8\times 8\) multiplier; a dot
indicates that the corresponding bit may be non-zero.
{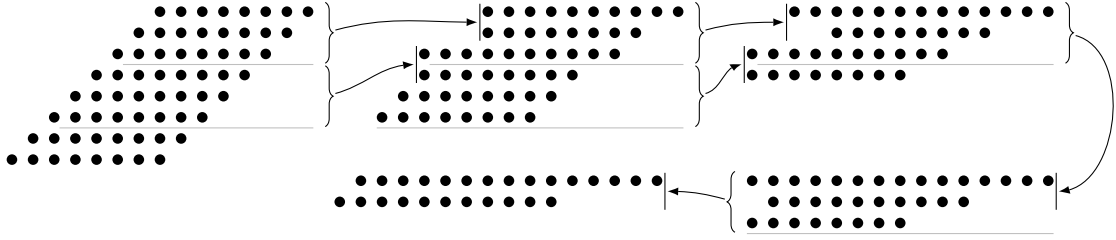
\begin{figure}[h]
  \centering
\begin{tikzpicture}
  [place/.style={radius=2pt,fill=black},
  rbrace/.style={decorate,decoration={mirror,brace,amplitude=3pt},xshift=3pt,yshift=0pt},
  lbrace/.style={ decorate,decoration={mirror,brace,amplitude=3pt},xshift=-3pt,yshift=0pt },
  myarrow/.style={arrows={-Latex[length=4pt,bend,line width=0pt]}}]
  \begin{scope}
    \foreach \j in {0,...,7}
    \foreach \i in {0,...,7}
    \fill (\i * 8pt + \j * 8pt, \j * 8pt) circle [place] {};
    \draw [rbrace] (14 * 8pt + 3.5pt, 5 * 8pt - 3.5pt) -- (14 * 8pt + 3.5pt, 7 * 8pt + 3.5pt)  node(c1) [midway] {};
    \draw [rbrace] (14 * 8pt + 3.5pt, 2 * 8pt - 3.5pt) -- (14 * 8pt + 3.5pt, 4 * 8pt + 3.5pt)  node(c2) [midway] {};
    \draw[myarrow, out=15,in=180] let \p1 = (c1) in (c1) to (140pt + 5*8 - 4pt , 7*8 - 4pt);
    \draw[myarrow, out=15,in=180] let \p1 = (c2) in (c2) to (140pt + 2*8 - 4pt , 5*8 - 4pt);
    \draw[gray!50] (5*8+2pt, 4*8+4pt) -- (14*8+2pt, 4*8+4pt);  
    \draw[gray!50] (2*8+2pt, 1*8+4pt) -- (14*8+2pt, 1*8+4pt);  
    \draw (140pt + 5*8 - 3pt , 7*8 + 3pt) to (140pt + 5*8 - 3pt , 6*8-3pt);
    \draw (140pt + 2*8 - 3pt , 5*8 + 3pt) to (140pt + 2*8 - 3pt , 4*8-3pt);
  \end{scope}
  \begin{scope}[xshift=140pt]
    \foreach \i in {5,...,14} \fill ( \i * 8pt, 56pt) circle [place] {};
    \foreach \i in {5,...,12} \fill ( \i * 8pt, 48pt) circle [place] {};
    \foreach \i in {2,...,11} \fill ( \i * 8pt, 40pt) circle [place] {};
    \foreach \i in {2,...,9} \fill ( \i * 8pt, 32pt) circle [place] {};
    \foreach \i in {1,...,8} \fill ( \i * 8pt, 24pt) circle [place] {};
    \foreach \i in {0,...,7} \fill ( \i * 8pt, 16pt) circle [place] {};
    \draw [rbrace] (14 * 8pt + 3.5pt, 5 * 8pt - 3.5pt) -- (14 * 8pt + 3.5pt, 7 * 8pt + 3.5pt)  node(c1) [midway] {};
    \draw [rbrace] (14 * 8pt + 3.5pt, 2 * 8pt - 3.5pt) -- (14 * 8pt + 3.5pt, 4 * 8pt + 3.5pt)  node(c2) [midway] {};
    \draw[myarrow, out=15,in=180] let \p1 = (c1) in (c1) to (140pt + 2*8 - 4pt , 7*8 - 4pt);
    \draw[myarrow, out=15,in=180] let \p1 = (c2) in (c2) to (140pt + 0*8 - 4pt , 5*8 - 4pt);
    \draw[gray!50] (2*8+2pt, 4*8+4pt) -- (14*8+2pt, 4*8+4pt);  
    \draw[gray!50] (-2pt, 1*8+4pt) -- (14*8+2pt, 1*8+4pt);  
    \draw (140pt + 2*8 - 3pt , 7*8 + 3pt) to (140pt + 2*8 - 3pt , 6*8-3pt);
    \draw (140pt + 0*8 - 3pt , 5*8 + 3pt) to (140pt + 0*8 - 3pt , 4*8-3pt);
  \end{scope}
  \begin{scope}[xshift=280pt]
    \foreach \i in {2,...,14} \fill ( \i * 8pt, 56pt) circle [place] {};
    \foreach \i in {4,...,11} \fill ( \i * 8pt, 48pt) circle [place] {};
    \foreach \i in {0,...,9} \fill ( \i * 8pt, 40pt) circle [place] {};
    \foreach \i in {0,...,7} \fill ( \i * 8pt, 32pt) circle [place] {};
    \draw [rbrace] (14 * 8pt + 3.5pt, 5 * 8pt - 3.5pt) -- (14 * 8pt + 3.5pt, 7 * 8pt + 3.5pt)  node(c1) [midway] {};
    \draw[myarrow, out=-15,in=0] let \p1 = (c1) in (c1) to (14*8 + 4pt , -8 -4pt);
    \draw[gray!50] (0*8+2pt, 4*8+4pt) -- (14*8+2pt, 4*8+4pt);
    \draw (14*8 + 3pt , 7*8 + 3-64pt) to (14*8 + 3pt , 6*8-3-64pt);
  \end{scope}
  \begin{scope}[xshift=280pt, yshift=-64pt]
    \foreach \i in {0,...,14} \fill ( \i * 8pt, 56pt) circle [place] {};
    \foreach \i in {1,...,10} \fill ( \i * 8pt, 48pt) circle [place] {};
    \foreach \i in {0,...,7} \fill ( \i * 8pt, 40pt) circle [place] {};
    \draw [lbrace] (- 3.5pt, 7 * 8pt + 3.5pt) -- (- 3.5pt, 5 * 8pt - 3.5pt) node(c1) [midway] {};
    \draw[myarrow, out=165,in=0] let \p1 = (c1) in (c1) to (-140pt + 14*8 - 4pt , 7*8 - 4pt);
    \draw[gray!50] (0*8-2pt, 4*8+4pt) -- (14*8+2pt, 4*8+4pt);  
    \draw (-148+14*8 + 3pt, 7*8 + 3pt) to (-148+14*8 + 3pt , 6*8-3pt);
  \end{scope}
  \begin{scope}[xshift=132pt, yshift=-64pt]
    \foreach \i in {0,...,14} \fill ( \i * 8pt, 56pt) circle [place] {};
    \foreach \i in {-1,...,9} \fill ( \i * 8pt, 48pt) circle [place] {};
  \end{scope}
\end{tikzpicture}
\caption{An \(8\times 8\) multiplier compression tree}
  \label{fig:8x8-bit-matrix}
\end{figure}}
We also split the verification task along the above separation in the
design. We define two separate RAC functions for integer multipliers
--- {\tt genPP} to generate the partial products and {\tt compress} to
mimic the compression tree and the final adder. For the final
correctness result, we need to prove that the sum of the partial
products generated by \texttt{genPP} is equal to the product, and that
the \texttt{compress} function's summation strategy is correct. In
this paper, we focus on automating the proofs of the implementations
of the compression tree, i.e., the RAC \texttt{compress} function.
Note that we verify the corresponding ACL2 definition of
\texttt{compress}, which is automatically generated by the RAC parser.
See examples below for both the RAC and its ACL2 translation for our
\(8\times 8\) running example, where some code is elided.

\begin{alltt}\small
// RAC Functions
ui16 compress(ui16 pp0, ui16 pp1, ui16 pp2, ... , ui16 pp7) \{
  ui16 l1pp0 = pp0^pp1^pp2;
  ui16 l1pp1 = ((pp0&pp1) | (pp0&pp2) | (pp1&pp2)) << 1;
  ...
  ui16 l4pp0 = l3pp0^l3pp1^l3pp2;
  ui16 l4pp1 = ((l3pp0&l3pp1) | (l3pp0&l3pp2) | (l3pp1&l3pp2)) << 1;
  return l4pp0 + l4pp1; \}
  
ui16 computeProd(ui8 a, ui8 b) \{
  array<ui16,8> pp = genPP(a, b);
  return compress(pp[0], pp[1], pp[2], pp[3], pp[4], pp[5], pp[6], pp[7]); \}

;; ACL2 Translation
(defund compress (pp0 pp1 pp2 pp3 pp4 pp5 pp6 pp7)
  (let* ((l1pp0 (setbits 0 16 15 0 (logxor pp0 pp1 pp2)))
         ...
         (l4pp0 (setbits 0 16 15 0 (logxor l3pp0 l3pp1 l3pp2)))
         (l4pp1 (setbits 0 16 15 0
                   (logior (logand l3pp0 l3pp1)
                           (logand l3pp0 l3pp2) (logand l3pp1 l3pp2)))))
    (bits (+ l4pp0 l4pp1) 15 0)))
\end{alltt}

Our new clause processor {\tt ctv-cp} may be invoked as follows to
automatically prove the correctness of {\tt compress}.
\begin{alltt}\small
(def-ctv-thm compress-lemma-8x8
  (implies (and (integerp pp0) (integerp pp1) (integerp pp2) (integerp pp3)
                (integerp pp4) (integerp pp5) (integerp pp6) (integerp pp7))
           (equal (compress pp0 pp1 pp2 pp3 pp4 pp5 pp6 pp7)
                  (bits (+ pp0 pp1 pp2 pp3 pp4 pp5 pp6 pp7) 15 0)))
  :expand (compress))
\end{alltt}

\section{Algorithm}

In principle, the correctness proof of the compression tree may be
developed by instantiating Theorem \ref{thm:add-3} from the RTL books
for each 3:2 compressor.

\begin{theorem}[Add-3]\label{thm:add-3} If \(x\), \(y\), and \(z\) are
  integers, and \(s = x\oplus y\oplus z\) and
  \(c = (x\wedge y)\vee(x\wedge z)\vee(y\wedge z)\), then
  \(s+2c = x + y + z\).
\end{theorem}

The clause processor {\tt ctv-cp} essentially does this instantiation
automatically. The high-level idea is simple; {\tt ctv-cp} works on
the LHS and RHS of a goal separately and processes terms on each side
into an internal format. It then applies a sequence of normalizing
transformations. At the end, if the resulting terms are the same, then
the goal is proven.

We describe the algorithm by considering the LHS of the conclusion of
{\tt compress-lemma-8x8} --- \texttt{(compress pp0 pp1 pp2 pp3 pp4 pp5
  pp6 pp7)}. First, {\tt ctv-cp} expands all the functions listed in
its \texttt{:expand} hint, i.e., \texttt{compress} in our example. The
untranslated body of this function contains a sequence of
let-bindings, whose translated version is a nested application of
lambda forms:
\begin{alltt}\small
((lambda (l0pp0 pp0 pp1 pp2 ... pp7)
  ...
    ((lambda (l4pp0 l4pp1)
       (bits (binary-+ l4pp0 l4pp1) '15 '0))
     l4pp0 (setbits '0 '16 '15 '0 (binary-logior ... ))) ...)
  (setbits '0 '16 '15 '0 (binary-logxor ... )) pp0 pp1 ... pp7)
\end{alltt}
The clause processor acts on this term by diving into the lambda
expressions to reach the inner-most term, \texttt{(bits (binary-+
  l4pp0 l4pp1) \textquotesingle15 \textquotesingle0)}. As it does so,
it also builds a \emph{substitution context} needed to interpret the
inner-most term. A \emph{substitution} is an association list mapping
symbols to ACL2 terms, and a substitution context is a list of such
substitutions. In our example, the first substitution is
\begin{alltt}\small
'((l0pp0 . (setbits '0 '16 '15 '0 (binary-logxor ... )))
  (pp0 . pp0) (pp1 . pp1) ... (pp7 . pp7)).
\end{alltt}
Once the inner-most expression is reached, the bit-width of the
expression is inferred (16 in the example), and the expression is
parsed into a data structure that represents its bitwise expansion.
This data structure is specified in BNF for brevity on the left side
below, but is defined using the FTY books~\cite{swords2015fix}. The right side
shows the interpretations for such data.
\newline

\begin{minipage}{0.3\textwidth}
\begin{alltt}\small
\(\mathit{bvfsl}\) := (cons \(\mathit{bvfs}\) \(\mathit{bvfsl}\))
      | nil
\(\mathit{bvfs}\) := '(\(\mathit{bvf}\) \(num\))            
\(\mathit{bvf}\)  := \(\mathit{bv}\)                          
      | '(:fas \(\mathit{bvf}\) \(\mathit{bvf}\) \(\mathit{bvf}\))   
      | '(:fac \(\mathit{bvf}\) \(\mathit{bvf}\) \(\mathit{bvf}\))   
                                          
                                          
\(\mathit{bv}\)   := '(:bit \(term\) \(num\))      
      | '(:v 0)                           
      | '(:v 1)
\end{alltt}
\end{minipage}
\hspace{0.5cm}\vline\hspace{0.5cm}
\begin{minipage}{0.5\textwidth}
\begin{alltt}\small
(cons \(a\) \(b\)) \(\mapsto\) (+ \((\mathit{interp}\ a)\) \((\mathit{interp}\ b)\))
 nil       \(\mapsto\) 0
'(\(a\) \(n\)) \(\mapsto\) (ash \((\mathit{interp}\ a)\) \(n\))

'(:fas \(a\) \(b\) \(c\)) \(\mapsto\) (logxor \((\mathit{interp}\ a)\) \((\mathit{interp}\ b)\) \((\mathit{interp}\ c)\))
'(:fac \(a\) \(b\) \(c\)) \(\mapsto\) (logior (logand \((\mathit{interp}\ a)\) \((\mathit{interp}\ b)\))
                   (logand \((\mathit{interp}\ a)\) \((\mathit{interp}\ c)\))
                   (logand \((\mathit{interp}\ b)\) \((\mathit{interp}\ c)\)))
'(:bit \(a\) \(n\)) \(\mapsto\) (bitn \((\mathit{interp}\ a)\) n)
'(:v 0)     \(\mapsto\) 0
'(:v 1)     \(\mapsto\) 1
\end{alltt}
\end{minipage}
\newline

For the running example, the bitwise expansion of the inner-most term is
\begin{alltt}\small
'(((:bit l4pp0 0) 0) ((:bit l4pp0 1) 1) ... ((:bit l4pp0 15) 15)
  ((:bit l4pp1 0) 0) ((:bit l4pp1 1) 1) ... ((:bit l4pp1 15) 15))
\end{alltt}
and its immediate interpretation (in untranslated form for
readability) is
\begin{alltt}\small
(+ (ash (bitn l4pp0 0) 0) (ash (bitn l4pp0 1) 1) ... (ash (bitn l4pp0 15) 15)
   (ash (bitn l4pp1 0) 0) (ash (bitn l4pp1 1) 1) ... (ash (bitn l4pp1 15) 15))
\end{alltt} 
\texttt{ctv-cp} generates the bitwise expansion by repeatedly calling
a function called \texttt{get-nth-bit}. When given a term \(x\) and a
bit position \(n\), this function outputs a \emph{bvf} form that has
the interpretation \texttt{(bitn \(x\) \(n\))}. The function
\texttt{get-nth-bit} knows how to parse some RTL library functions
such as \texttt{bits}, \texttt{setbits}, etc., that appear in code
generated by the RAC parser. It can also recognize expressions
emerging from instances of 3:2 compressors and generate \textit{bvf}
forms of type \texttt{:fas} or \texttt{:fac}. Specifically, a term of
the form \texttt{(logxor \(a\) \(b\) \(c\))} yields
\texttt{(:fas \(a'\) \(b'\) \(c'\))}, and a term of the form
\texttt{(logior (logand \(a\) \(b\)) (logand \(a\)
  \(c\)) (logand \(b\) \(c\)))} gives the output \texttt{(:fac \(a'\)
  \(b'\) \(c'\))}, where \(a'\), \(b'\) and \(c'\) are bvf's obtained by
recursively calling \texttt{get-nth-bit} on \(a\), \(b\), and \(c\)
respectively. Note that if \texttt{get-nth-bit} fails to parse a term,
then it--and consequently \texttt{ctv-cp}--aborts with an error.

After parsing, {\tt ctv-cp} applies the following transformations
until the substitution context is empty.
\begin{enumerate}
\item Match all bvfs of the form \texttt{((:fas \(a\) \(b\) \(c\))
    \(k\))} and \texttt{((:fac \(a\) \(b\) \(c\)) \(k+1\))}, and
  replace them with the three bvfs' \texttt{(\(a\) \(k\))},
  \texttt{(\(b\) \(k\))}, and \texttt{(\(c\) \(k\))}.
\item Apply the most recent substitution in the context to get a new
  \(\mathit{bvfsl}\).
\end{enumerate}
The first transformation is valid because of the \texttt{add-3} lemma.
To optimize the matching algorithm, we normalize and sort the
\textit{bvfs} terms. The function \texttt{get-nth-bit} is again used
by the substitution step --- substituting \texttt{(\(x\) .\ \(term\))}
in the \textit{bv} form \texttt{(:bit \(x\) \(l\))} gives
\texttt{(get-nth-bit \(term\) \(l\))}.

An important detail is that the transformations are justified by
lemmas in the RTL books that have \texttt{integerp} type constraints;
see, e.g., the \texttt{add-3} lemma. We defer discharging these
hypotheses until the end. All \texttt{ctv-cp} functions maintain a
list of terms that need to satisfy \texttt{integerp}, and syntactic
analysis is done to resolve such hypotheses whenever a substitution is
made. If the final transformed terms for LHS and RHS match, the clause
processor tries to prove these type hypotheses under the original
assumptions of the theorem; if it cannot, then it prompts the user to
supply any missing assumptions.

\section{Observations and Related Work}
The largest multipliers that we have used \texttt{ctv-cp} on so far at
Arm have \(64\times 64\)-bit Dadda and Wallace compression trees; the
runtime is less than 1 second. The automation and speed of
\texttt{ctv-cp} reduces the ACL2 proof development effort for integer
multipliers and facilitates quick equivalence checks because the RAC
models can faithfully replicate the RTL. We refrain from doing a
formal complexity analysis for \texttt{ctv-cp}, but note that its
runtime is proportional to the size of the \(\mathit{bvfsl}\) terms and the
number of substitutions in the design. The size of the terms is never
larger than the product of the number of the initial partial products
and the multiplication size (i.e., 16 for an \(8\times 8\)-bit
multiplier). Thus, we expect \texttt{ctv-cp} to scale for the
multipliers we deal with at Arm.

An alternative approach for verifying compression trees would be to
apply rewriting after the beta-reduction of lambda terms. For
efficiency, such an approach would need structure sharing using
hash-consing, outside-in rewriting, and optimized algorithms for term
matching. Our implementation is simple; it operates on lambda terms
and applies the matching algorithm from the inner-most term outwards
before applying substitutions; this is equivalent in principle to the
alternative approach above, and obviates the need for such nontrivial
optimization techniques.

In related work~\cite{Temeltacas24,TemelACL2-2022}, the author
develops an efficient, automatic tool, VeSCMul, for end-to-end proofs
of a wide variety of multiplier designs in ACL2. A rewriting-based
approach is used that employs optimization techniques to avoid costly
backchaining. Unfortunately, VeSCMul does not currently work with
functions in the RTL books, which are present in the code generated by
the RAC parser. Instead of implementing a translator, we developed
\texttt{ctv-cp} which has a simple implementation, works seamlessly
with our existing verification methodology, and has the advantage that
it normalizes terms until fixpoint, which is conducive to producing
informative messages if any errors are encountered.

In the future, we plan to develop automation for reasoning about the
partial product generation step to reduce the verification overhead of
obtaining end-to-end correctness proofs for integer multipliers and
subsequently, other design units that include them.

\bibliographystyle{eptcs}
\bibliography{generic}

\end{document}